\documentclass[reprint,aps, superscriptaddress,nofootinbib]{revtex4-1}

\usepackage{graphicx}
\usepackage{dcolumn}
\usepackage{bm}
\usepackage{amsmath}
\usepackage{amssymb}
\usepackage{amsfonts}
\usepackage{amsthm}
\usepackage{color}
\usepackage{subfig}
\usepackage{MnSymbol}
\begin{document}

\title{Topological defect formation in a phase transition with tunable order}

\date{\today}

\author{Fumika Suzuki}
\email{fsuzuki@lanl.gov}
 \affiliation{%
Theoretical Division, Los Alamos National Laboratory, Los Alamos, New Mexico 87545, USA
}

 \affiliation{%
Center for Nonlinear Studies, Los Alamos National Laboratory, Los Alamos, New Mexico 87545, USA
}

\author{Wojciech H. Zurek}

 \affiliation{%
Theoretical Division, Los Alamos National Laboratory, Los Alamos, New Mexico 87545, USA
}

  \begin{abstract}
The Kibble-Zurek mechanism (KZM) describes the non-equilibrium dynamics and topological defect formation in systems undergoing second-order phase transitions. KZM has found  applications in fields such as cosmology and condensed matter physics. However, it is generally not suitable for describing first-order phase transitions. It has been demonstrated that transitions in systems like superconductors or charged superfluids, typically classified as second-order, can exhibit weakly first-order characteristics when the influence of fluctuations is taken into account. Moreover, the order of the phase transition (i.e., the extent to which it becomes first rather than second order) can be tuned. We explore quench-induced formation of topological defects in such tunable phase transitions and propose that their density can be predicted by combining KZM with nucleation theory.\end{abstract}

\maketitle


\begin{figure*}
\includegraphics[clip,width=2\columnwidth]{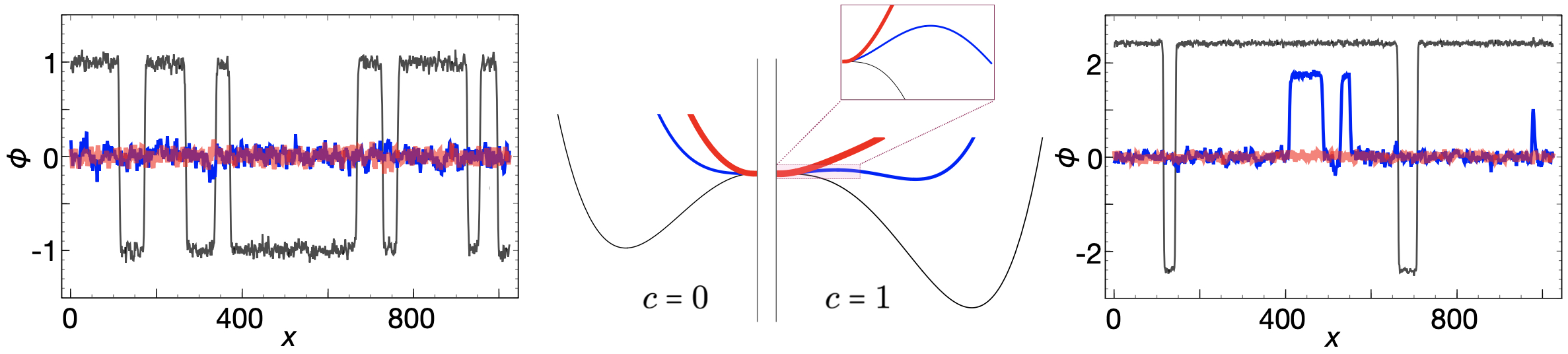} 
\caption{Snapshots of $\phi$ and  corresponding $V (\phi)$  following  second-order phase transition with $c=0$  (left) and  phase transition with $c=1$ (right). Plots of $\phi(x)$ at different stages of the quench, starting with $\epsilon =-1.5$ (thick red line), $\epsilon=-0.45$ (blue line), and $\epsilon =1$ (thin black line) are shown.} 
\label{fig1}
\end{figure*}

\begin{figure*}
{%
\includegraphics[clip,width=2\columnwidth]{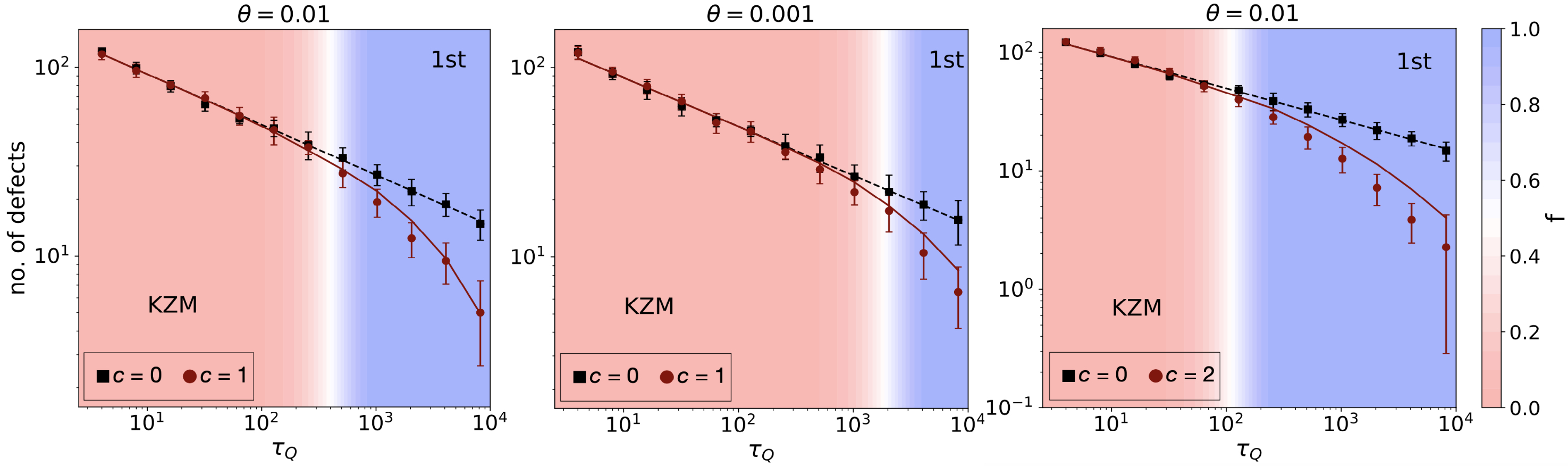} 
}
\caption{Number of defects as a function of quench timescale $\tau_Q$. Black squares represent numerical results for  $c=0$ where KZM is expected to hold. The dashed black  line represents the best fit of the black squares. Dark red circles represent numerical results for $c=1$, $\theta=0.01$ (left),  for $c=1$, $\theta=0.001$ (middle), and for  $c=2$, $\theta=0.01$ (right) respectively. The solid dark red  lines represent the number of defects derived from Eq. (\ref{formula}). The fraction of space $\mathfrak{f}$ occupied by the new phase due to nucleation events   (\ref{fraction}) is depicted using a color plot.  } 
\label{fig2}
\end{figure*}

The Kibble-Zurek mechanism (KZM) combines Kibble's observation of the inevitability of topological defect formation in cosmological phase transitions \cite{kibble,kibble2} with the theory proposed by one of us  \cite{whz,whz2,whz3} that relates their density to the critical slowing down and, hence, to the universality class of the second-order phase transition. The resulting KZM predicts defect density as a function of the quench rate during second-order phase transitions, in both classical and quantum settings \cite{kzm,kzm2,Damski,kzm3,Dziarmaga,Dziarmaga2, Polkovnikov, Polkovnikov2, Sadhukhan, Francuz, kzm4,kzm5,kzm6,kzm7,kzm8,kzm9,kzm10,kzm11,kzm12,kzm13, kzm14,kzm15}. 
It finds applications in 
condensed matter physics \cite{whz, whz2, whz3}, 
cosmological phase transitions \cite{kibble, kibble2,cos, cos2, cos3}, superconductors \cite{sup}, liquid crystals \cite{liquid, liquid2}, superfluids \cite{fluid,fluid2,fluid3}, ultracold chemistry \cite{kzm15}, Bose-Einstein condensates \cite{bec, bec2, bec3,bec4} and quantum computing \cite{qc,exp2}. 

However, KZM is generally not suitable for describing first-order phase transitions. In \cite{weakly}, it has been demonstrated that the transitions associated with superconductors or 
superfluids can exhibit weakly first-order characteristics \cite{weakly2}. This should allow one to tune the order of the transition between the second and first order, with the weakly first-order characteristics inbetween. Given the critical properties shared between, e.g., smectic-$A$ liquid crystals and superconductors, the transitions in liquid crystals can also exhibit a weakly first-order nature. In particular, there is now compelling evidence that the order of the Fredericks phase transition can be `tuned' in this manner \cite{Fredericks}. 

In this paper, we demonstrate that the formation of topological defects in those systems can resemble either a second-order or a first-order phase transition, or fall  inbetween these two regimes (i.e., become weakly first-order). This variation depends on factors such as the strength of the first-order component in the free energy, the quench timescale, and the temperature. 

While KZM has been investigated numerically \cite{numeric, numeric2, numeric3, lg} and experimentally \cite{sup,liquid,liquid2,fluid,fluid2,fluid3,exp, qc, exp2, exp3, exp4, exp5}, its  applicability to weakly first-order or tunable phase transitions is an open question. The following analysis demonstrates that  KZM can remain viable for predicting the density of defects generated in a phase transition with tunable order when it is integrated with thermally activated nucleation \cite{nucleation, nucleation2, nucleation3, nucleation5, nucleation4}. 

We note that Kibble suggested initially \cite{kibble} that thermal activation determines the density of defects even in the second order non-equilibrium phase transitions \cite{TK}. In contrast to KZM, thermal activation would result in defect densities independent of the quench rate. Nevertheless, as we shall see, thermally activated nucleation can compete with KZM in determining the density of topological defects in the tunable transitions we consider.    

Here, we first present numerical results illustrating the interplay of critical slowing down and thermally activated nucleation in the formation of topological defects in a phase transition with tunable order. We then provide an analytical interpretation of the results.\\

To explore KZM in a  phase transition with tunable order, we examine the numerical evolution of a one-dimensional system governed by the equation of motion for a real scalar field $\phi$. The equation is derived from the modified Landau-Ginzburg potential,
\begin{eqnarray}\label{pot}
V(\phi)= (\phi^4-2\epsilon \phi^2 )/8-c|\phi|^3/3
\end{eqnarray}
where  the first two terms account for the typical second-order phase transition behavior, whereas the third term introduces the first-order characteristics as presented in \cite{weakly} (see Fig. 1). Here, the constant $c$ represents  the strength of the term responsible for first-order nature of the phase transition.

We assume that $\epsilon $ follows a linear quench, $\epsilon (t)=t/\tau_Q$ with $\tau_Q$ representing the quench timescale.

The system is in contact with a thermal reservoir and it obeys the Langevin equation,
\begin{eqnarray}\label{eq}
\ddot{\phi}+\eta \dot{\phi} -\partial_{xx}\phi +\partial_{\phi} V(\phi)=\vartheta (x,t)
\end{eqnarray}
where the noise term $\vartheta$ has correlation properties,
\begin{eqnarray}
\langle \vartheta (x,t), \vartheta  (x',t')\rangle = 2\eta \theta \delta (x'-x)\delta (t'-t)
\end{eqnarray}
with the temperature of the reservoir $\theta$ and $\eta$ is the overall damping constant. In this paper, we set $\eta=1$.

When $c=0$, we recover the ordinary second-order phase transition where $\epsilon$ measures the distance from the critical point. $t<0$ and $t>0$ represent the time before and after  the transition at $\epsilon=0$ respectively. This scenario was throughly investigated in \cite{lg}.

When $c > 0$, a characteristic of a   first-order phase transition  emerges: For $\epsilon < -c^2$, the potential exhibits symmetry with a single minimum, similarly to a second-order phase transition. However, for $-c^2 <\epsilon <0$, it develops two minima at $\phi =\pm(c+\sqrt{c^2+\epsilon})$ corresponding to the new phase in addition to the existing one at $\phi=0$ representing the old phase, leading to nucleation associated with the first-order phase transition. The positions of these nucleation barrier peaks are  $\pm \phi_{\rm barrier}$ where $ \phi_{\rm barrier} = c-\sqrt{c^2+\epsilon}$, and their height is  
\begin{eqnarray}
h_{\rm barrier}=-\frac{1}{24}(c-\sqrt{c^2+\epsilon})^2(3\epsilon +2c(c-\sqrt{c^2+\epsilon})).
\end{eqnarray}
This indicates that the positions $\phi_{\rm barrier}$ and the height $h_{\rm barrier}$ of the nucleation barriers approach $0$ as $\epsilon \rightarrow 0$. Hence, with both $c=0$ and  $c>0$, the potential with two minima eventually emerges for $\epsilon > 0$. Fig. \ref{fig1} illustrates the snapshots of $\phi$ and  corresponding $V(\phi)$ following  second-order phase transition (left) and phase transition with non-zero $c$ (right). For the second-order phase transition, $\phi$ initially fluctuates around its expectation value $\langle \phi\rangle =0$ when $\epsilon<0$. After the symmetry breaking takes place (i.e., $\epsilon >0$), $\phi$ is forced to choose one of two minima and gradually settles locally around $\langle \phi \rangle \approx \pm \sqrt{\epsilon}$ while forming defects. For a  phase transition with $c>0$, $\phi$ follows the similar transition, except that nucleations can occur when $-c^2 < \epsilon < 0$. Our primary interest lies on assessing the impact of these nucleation events on the density of defects  after the  transition  ($\epsilon >0$).

Following the method described in \cite{lg}, we numerically investigate the number of defects generated by phase transitions as a function of the quench timescale $\tau_Q$. We initiate the time evolution obeying Eq. (\ref{eq}) with $\epsilon = -2$  and conclude it when $\epsilon$ reaches 5. The number of defects is determined by counting the points where $\phi = 0$ at $\epsilon = 5$.  We performed 15 numerical simulations of the phase transition for each $\tau_Q$ and obtained   Fig. \ref{fig2}.
 Black squares and dark red circles represent numerical results for $c = 0$ (purely second-order phase transition) and $c >0 $ respectively. The dashed black  line represents the best fit of the black squares. The best fit corresponds to $n_{KZM}\propto \tau_{Q}^{-a}$ where $a=0.267\pm 0.029$ which agrees closely with the theoretical prediction of KZM, $a=1/4$ \cite{whz,whz2,whz3}. As the quench timescale $\tau_Q$ increases, we notice a  pronounced deviation of the dark red circles from the prediction of KZM depicted by the dashed black line. This departure can be attributed to the increased likelihood of nucleation events. In the middle panel, we have a decrease in the nucleation rate due to a low temperature  $\theta=0.001$.   Because of the low nucleation rate, there is only a small overall deviation from the predictions of KZM. On the other hand, the right panel presents the results for larger value of $c=2$, indicating a stronger first-order phase transition term in the  potential (\ref{pot}). In this case,  a significant departure from KZM is observed even when $\tau_Q$ is relatively small. 
 
 These plots can be understood as follows: When $\phi$ fluctuates around its expectation value $\langle \phi \rangle =0$ initially, it starts to interact with the nucleation barriers at $t=t_1$ when $\sqrt{\langle \phi^2\rangle}$ is approximately equal to the location of the  barriers $\phi_{\rm barrier}$.  In the vicinity of $\phi=0$, the potential can be approximated by a harmonic potential $V_{\rm har}(\phi)=\frac12 \omega^2 \phi^2$ where $\omega$ is given by $V_{\rm har}(\phi_{\rm barrier})=h_{\rm barrier}$. Since the temperature $\theta$ corresponds to the energy of $\phi$, we have $\langle \phi^2 \rangle \approx \theta/\omega^2$.  Therefore, $\phi$ starts to interact with the barrier at $t_1$ when $(\theta/\omega^2)^{1/2}=\phi_{\rm barrier}$. After $t_1$, there is a possibility of nucleation occurring. 
 
 The nucleation rate per unit length of the metastable state around $\phi=0$ is  given by \cite{nucleation}
\begin{eqnarray}\label{gamma}
\Gamma (\epsilon(t)) =A\exp [-B(\epsilon(t))/\theta]
\end{eqnarray}
where
\begin{eqnarray}\label{B}
B(\epsilon(t))=2\int_0^{\phi_{TP}}d\phi \sqrt{2V(\phi)}.
\end{eqnarray}
Here $\phi_{TP}$ is the classical turning point such that $V(\phi=0)=V(\phi_{TP})=0$. The prefactor $A$ exhibits only a soft dependence on temperature and $\epsilon$. In this paper, we   set $A\approx 0.4$ obtained numerically. Since $\epsilon$ is time-dependent, $B$ and $\Gamma$ are time-dependent. In particular, $B\rightarrow 0$ as $\epsilon \rightarrow 0$. This suggests that as the parameter $\epsilon$ approaches $0$, the influence of the barriers becomes insignificant in comparison to the kinetic energy of the field $\langle\dot{\phi}^2\rangle$, given by the temperature $\theta$. Then $\phi$ undergoes a transition to one of the two broken symmetry minima, much like what occurs in a second-order phase transition. In our model, this behavior is observed after time $t_2$ when the energy of the nucleation barriers becomes equal to the kinetic energy of $\phi$, $B (\epsilon(t_2))=\theta$ (supplemental material \cite{supp}).

The  fraction of space $\mathfrak{f}$ occupied by the new phase due to nucleation events  during the period between $t_1$ and $t_2$ can be obtained using the Avrami equation \cite{av4, av3, av,av2, av5}. This equation describes the progress of phase transformations via a nucleation-growth process in first-order phase transitions, under the assumption that the transformation follows  a  sigmoidal function. It is applied in various areas, including cosmology, to describe the fraction of space that has undergone transition from the false vacuum to the true vacuum \cite{false,false2,false3}. It can be derived by assessing the probability that a particular point in space is not enclosed within any true vacuum bubbles. In supplemental material \cite{supp}, we provide a brief derivation of the equation by following \cite{false2,false3}. The equation reads
\begin{eqnarray}\label{fraction}
\mathfrak{f}= 1- \exp (-\Omega )
\end{eqnarray}
where 
\begin{eqnarray}\label{fraction2}
\Omega=\int_{t_1}^{t_2}\Gamma (\epsilon(t)) \mathcal{V}(t, t_2) dt.
\end{eqnarray}
Here $\mathcal{V}(t,t_2)= \int_t^{t_2}v(\epsilon(\tau))d\tau$ represents the volume of a nucleated bubble at the time $t_2$, which was formed at time $t$.  $v$ is the bubble wall velocity. $\mathfrak{f}$ describes the  fraction of space transformed to the new phase between $t_1$ and $t_2$, and $\mathfrak{f}=1$ when  the entire space is covered by the new phase through a nucleation-growth process during the time interval. The velocity $v$ is dependent on $\theta$, $\eta$, and $\epsilon$. Since we  fix $\theta$ and $\eta$ during the time evolution in our model, we only analyze the $\epsilon$-dependence of $v$ by numerical simulations as follows. 

When $\epsilon$ with $-c^2<\epsilon<0$ is held fixed and time-independent, both the nucleation rate $\Gamma$ and the velocity $v$ become time-independent, and only a nucleation-growth process takes place. Then we have the general Avrami equation in one dimension written  as
\begin{eqnarray}
\mathfrak{f}_{\rm fixed}=1-\exp \left(-\frac12  v (\epsilon_{\rm fixed}) \Gamma (\epsilon_{\rm fixed})  t^2\right).
\end{eqnarray}
By fitting this equation to the numerical results of the time evolution of  the   fraction of space occupied by the new phase at each fixed $\epsilon$, we  obtain the $\epsilon$-dependence of $ v$, which will then be substituted into  Eqs. (\ref{fraction}), (\ref{fraction2}) in the following discussion (supplemental material \cite{supp}). For $c=1$ and the temperature $\theta=0.01$, $v(\epsilon) = 0.026\epsilon + 0.016$ was obtained by the method described above. For different values of $c$ and $\theta$, we repeated the same procedure to obtain corresponding $v$.

In the absence of nucleation events, the  fraction of space occupied by the new phase due to nucleations is zero, i.e., $\mathfrak{f}=0$. The field $\phi$ then  would follow  second-order phase transition behavior obeying KZM  even in the presence of a non-zero $c$. Conversely, with an increase in $\mathfrak{f}$, the behavior of the first-order phase transition  becomes dominant. It can be assumed that the density of defects for the  fraction of space $\mathfrak{f}$ follows nucleation theory, while the density for the remaining space $(1-\mathfrak{f})$ obeys KZM. The number of defects generated in a phase transition with  tunable order can then be estimated as
\begin{eqnarray}\label{formula}
n= \mathfrak{f} n_{nuc} +(1-\mathfrak{f}) n_{KZM}
\end{eqnarray}
where $n_{KZM}$ obeys  KZM in the second-order phase transition, i.e., $n_{KZM}\propto \tau_Q^{-a}$ with $a=1/4$ in our model.  In general, the number of defects $n_{nuc}$ generated by a nucleation-growth process increases with the increase in the nucleation rate $\Gamma$ and decreases with the rise of the bubble wall velocity $v$. It is because a larger value of $\Gamma$ leads to the growth of the number density of bubbles. Consequently, the distance between bubbles shortens, and the average time before a collision between two domains decreases. Conversely, the domain size increases for larger $v$, as bubbles grow more rapidly  before the collision \cite{nucleation4}. Since $\Gamma$ is a complex function of $\epsilon$,  we numerically obtain the $\epsilon$-dependence of $n_{nuc}$ by performing the nucleation process for each fixed $\epsilon$.   $n_{nuc}$ exhibits an almost linear dependence on $\epsilon$ within the relevant small $\epsilon$ range of interest (i.e., $-c^2 <\epsilon <0$). Numerically, we obtained the equation $n_{nuc} \approx 144\epsilon + 74$ for $c=1$ and $\theta=0.01$ (supplemental material \cite{supp}). For different values of $c$ and $\theta$, we repeated the same procedure to derive the equation for $n_{nuc}$. Since $\epsilon$ changes over time during the phase transition, $n_{nuc}$ also depends on time. We assume that the number of defects created by the nucleation-growth process throughout time evolution can be approximated by the time-averaged value, $n_{nuc} \sim n_{nuc}(\epsilon^{*})$ where $\epsilon^{*}$ represents the value of $\epsilon$ at the time when the  fraction of space occupied by the new phase reaches half of $\mathfrak{f}$, i.e., $\epsilon^{*}=\epsilon (t^{*})$ where $\mathfrak{f}^{*}=1-\exp\left(-\int_{t_1}^{t^{*}}\Gamma (\epsilon (t))\mathcal{V} (t, t^{*})dt\right) $ and $\mathfrak{f}^{*}=\mathfrak{f}/2$ (supplemental material \cite{supp}). After $t_2$, the nucleation barriers diminish  in comparison to the kinetic energy of $\phi$, leading a behavior  similar to a second-order transition. The number of defects generated within this regime obeys KZM and is given by $n_{KZM}$.

 The solid dark red  lines in Fig. \ref{fig2} correspond to the number of defects derived from Eq. (\ref{formula}).  They show reasonable agreement with the numerical results represented by the dark red circles. The  fraction of space $\mathfrak{f}$ occupied by the new phase due to nucleation events  (\ref{fraction}) is depicted using a color plot.  The phase transition occurs so rapidly that   $\phi$ does not have sufficient time to interact with the nucleation barriers  for the small  quench timescale $\tau_Q$. Consequently, nucleation does not occur, and the prediction of KZM remains valid in this regime.  As $\tau_Q$ increases, $\mathfrak{f}$ also grows, and  we see the transition into a regime where the behavior of the  first-order phase transition becomes dominant, leading to a significant departure from KZM. As the temperature $\theta$ decreases, the nucleation rate also decreases, which in turn supports the applicability of KZM for even larger values of  $\tau_Q$. Conversely, with larger values of $c$, the nucleation barriers persist for a longer duration, resulting in  deviations from the predictions of KZM even with relatively small $\tau_Q$.

\begin{figure}
{%
\includegraphics[clip,width=1\columnwidth]{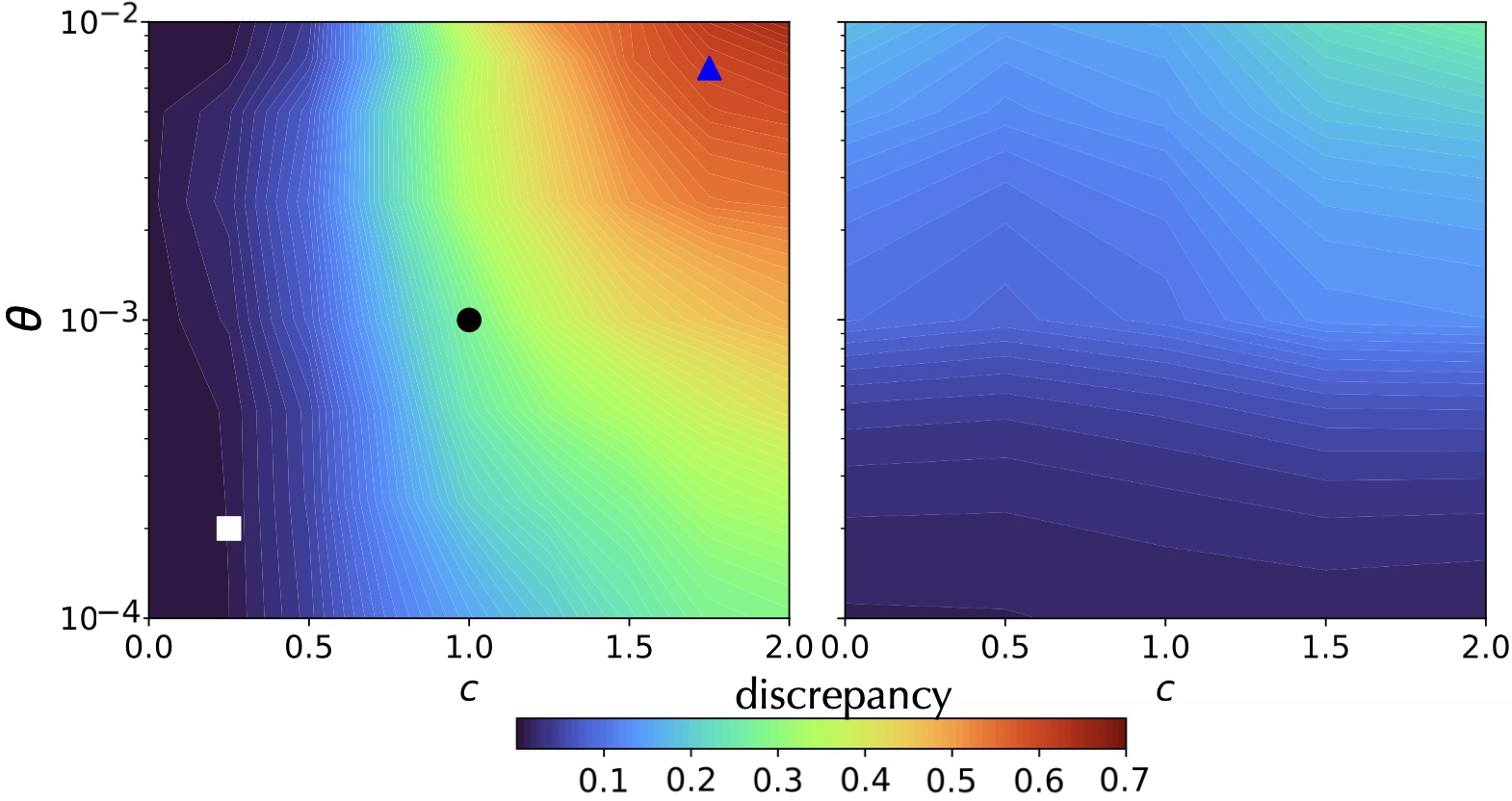} 
}
\caption{The discrepancy  between theory and numerical experiment as the function of $c$ and temperature $\theta$. The quench timescale $\tau_Q = 2048$. Left: the remaining discrepancy  between the numerical results and the predictions of KZM alone. Right: the  remaining discrepancy between the numerical results and the predictions of Eq. (\ref{formula}) that combines KZM and nucleation.} 
\label{fig3}
\end{figure}

By modifying Eq. (\ref{formula}), we  can estimate the discrepancy $\delta$ between the value predicted by KZM and the  value obtained in a numerical simulation of a phase transition with tunable order,
\begin{eqnarray}\label{dev}
\delta =\left|\frac{n-n_{KZM}}{n_{KZM}}\right|=\mathfrak{f}\left|1-\frac{n_{nuc}}{n_{KZM}}\right|.
\end{eqnarray}

Fig. \ref{fig3} shows the discrepancy as the function of $c$ and temperature $\theta$ where the  quench timescale $\tau_Q = 2048$. The left panel shows the discrepancy between the numerical results and the predictions of KZM. As $c$ and $\theta$ increase, we observe a larger discrepancy, represented by the red color. For higher temperature $\theta$, the kinetic energy of $\phi$ increases, thereby enhancing the likelihood of nucleation occurring prior to the second-order phase transition. For larger $c$, the nucleation barriers persist longer,  thus increasing the occurrence of nucleation events. The  fraction of space occupied by the new phase that forms due to nucleation events and the discrepancy can be evaluated using Eq. (\ref{fraction}) and Eq. (\ref{dev}). This yields $\mathfrak{f}=0$, $\delta=0$ for $c=0.25$, $\theta=0.0002$ (white square), $\mathfrak{f}=0.55$, $\delta=0.16$ for $c=1$, $\theta=0.001$ (black circle), and $\mathfrak{f}=1$, $\delta=0.66$ for $c=1.75$, $\theta=0.007$ (blue triangle).  These values  closely align with the numerical results illustrated in the figure. The right panel displays the discrepancy between the numerical results and the predictions of Eq. (\ref{formula}), i.e., $|(n_{\rm numeric}-n)/n|$, where $n_{\rm numeric}$ represents the mean value of the number of defects obtained numerically and $n$ is given by Eq. (\ref{formula}). It demonstrates that Eq. (\ref{formula}) effectively predicts the number of defects generated in a phase transition with tunable order.\\

In this paper, we investigated topological defect formation in a phase transition with tunable order. Such phase transitions can be observed in various systems, including superconductors, charged superfluids and liquid crystals. It has been shown that KZM can remain effective in predicting defect density  when integrated with nucleation theory.

The  fraction  of space $\mathfrak{f}$ occupied by the new phase due to nucleation events  from the Avrami equation proves to be useful for distinguishing between regimes governed by KZM and those dominated by nucleation processes.  When $\mathfrak{f} = 0$, nucleations do not occur prior to the second-order phase transition, and  KZM can provide an accurate prediction of defect density. When $\mathfrak{f} = 1$, the entire space undergoes a transition to the new phase through nucleation-growth processes before the second-order phase transition, and the defect density is determined by nucleation theory. When $0 < \mathfrak{f} < 1$, we  postulated  that the defect density in the region covered by the new phase between times $t_1$ and $t_2$ can be described by nucleation theory, while the  density in the remaining space follows  KZM. 

Our numerical results provide support for this conjecture. It is conceivable to validate our  findings within the phase transitions of liquid crystals \cite{Fredericks} in the future.

\begin{acknowledgements}
We thank Pablo Laguna for helpful discussions. F.S. acknowledges support from the Los Alamos National Laboratory LDRD program under project number 20230049DR and the Center for Nonlinear Studies under project number 20220546CR-NLS.
\end{acknowledgements}

\onecolumngrid

\appendix

\renewcommand{\thefigure}{S\arabic{figure}}
\setcounter{figure}{0}
\renewcommand{\thetable}{S\arabic{table}}
\setcounter{table}{0}
\setcounter{equation}{0}
\renewcommand{\theequation}{S.\arabic{equation}}

\section*{SUPPLEMENTAL MATERIAL}

\subsection*{Equations to obtain $t_1$ and $t_2$}

Near $\phi=0$, the potential can be approximated by a harmonic potential $V_{\rm har}(\phi)=\frac12 \omega^2 \phi^2$ where $\omega$ is given by $V_{\rm har}(\phi_{\rm barrier})=h_{\rm barrier}$. Since  $\phi_{\rm barrier} = c-\sqrt{c^2+\epsilon}$ and
\begin{eqnarray}
h_{\rm barrier}=-\frac{1}{24}(c-\sqrt{c^2+\epsilon})^2(3\epsilon +2c(c-\sqrt{c^2+\epsilon})),
\end{eqnarray}
we obtain
\begin{eqnarray}
\omega(\epsilon) = \frac{\sqrt{8c^4 + 12c^2\epsilon + 3\epsilon^2 - 8c^3\sqrt{c^2 + \epsilon} - 8c\epsilon\sqrt{c^2 + \epsilon}}}{2\sqrt{-3\epsilon + 6c(-c + \sqrt{c^2 + \epsilon})}}.
\end{eqnarray}

Then $t_1$ can be obtained from $\epsilon$ which satisfies
\begin{eqnarray}
\sqrt{\langle \phi^2\rangle} \approx (\theta/\omega^2)^{1/2}=\phi_{\rm barrier}.
\end{eqnarray}

$t_2$ represents the time when $B(\epsilon(t_2)) = \theta$. $B$ can be analytically evaluated as
\begin{eqnarray}
B(\epsilon(t))&=&2\int_0^{\phi_{TP}}d\phi \sqrt{2V(\phi)}\nonumber\\
&=&\frac{2}{27}\Biggr(3\sqrt{-2\epsilon}(4c^2+3\epsilon)-2(8c^3+9c\epsilon)\log(4c+3\sqrt{-2\epsilon})-\frac{c(8c^2+9\epsilon)^{5/4}(\sqrt{8c^2+9\epsilon}-2\sqrt{2}c) \log (16c^2+18\epsilon)}{\sqrt{16c^2\sqrt{8c^2+9\epsilon}+9\epsilon\sqrt{8c^2+9\epsilon}-32\sqrt{2}c^3-36\sqrt{2}c\epsilon}}\Biggr).\nonumber\\
\end{eqnarray}

\subsection*{The bubble wall velocity $v$ from the Avrami equation}

The Avrami equation, also known as the Johnson–Mehl–Avrami–Kolmogorov (JMAK) equation \cite{av4,av3,av,av2,av5}, was originally developed to characterize phase transformations involving nucleation-growth processes, such as crystallization. In cosmology, the equation is used to describe  the fraction of space that is covered by true vacuum bubbles and that remains in the false vacuum. It is typically derived by assessing the probability that a specific point in space is not enclosed within any true vacuum bubbles as follows \cite{false2, false3}. Let $\rho(V)dV$ be the density of bubbles with volume between $V$ and $V+dV$, and let $\mathfrak{g}(V_1, V_2)$ be the probability that a given point is not contained in any bubble of volume between $V_1$ and $V_2$. Then
\begin{eqnarray}
\mathfrak{g}(V_1,V_2+dV_2)=\mathfrak{g}(V_1,V_2)\mathfrak{g}(V_2,V_2+dV_2)=\mathfrak{g}(V_1,V_2)[1-\rho (V_2) V_2 dV_2].
\end{eqnarray}

This leads
\begin{eqnarray}
\frac{d \mathfrak{g}(V_1,V_2)}{dV_2}=\frac{\mathfrak{g} (V_1,V_2+dV_2)-\mathfrak{g}(V_1,V_2)}{dV_2}=-\rho (V_2) V_2 \mathfrak{g} (V_1,V_2)
\end{eqnarray}
and
\begin{eqnarray}
\mathfrak{g}(V_1,V_2)=\exp \left(-\int_{V_1}^{V_2} dV \rho (V) V\right).
\end{eqnarray}

The probability $\mathfrak{g}$ that a given point in space is not enclosed within any true vacuum bubbles is given by setting $V_1=0$ and $V_2=\infty$,
\begin{eqnarray}
\mathfrak{g}= \mathfrak{g}(0,\infty)=\exp \left(-\int_0^{\infty} dV \rho(V) V\right)
\end{eqnarray}
Here $\int_0^{\infty} dV \rho(V) V$ is the total volume in bubbles per unit volume of space. Since the total number of bubbles formed per unit time per unit volume is given by $\Gamma (t')$, and   the volume of a nucleated bubble at the time $t$, which was formed at time $t'$ is given by $\mathcal{V}(t',t)= \int_{t'}^{t}v(\epsilon(\tau))d\tau$ where  $v$ is the bubble wall velocity, it can be replaced by $\int_{0}^{t}\Gamma (\epsilon(t')) \mathcal{V}(t', t) dt'$.

Therefore the fraction of space remaining in the old phase  can be expressed as
\begin{eqnarray}
\mathfrak{g}=\exp\left(-\int_{0}^{t}\Gamma (\epsilon(t')) \mathcal{V}(t', t) dt'\right)
\end{eqnarray}
$\mathfrak{g}$ is often called  the false vacuum fraction in cosmology. Consequently, the fraction of space occupied by the new phase due to nucleations is given by $\mathfrak{f}=1-\mathfrak{g}$.

When we perform numerical simulations of the Langevin equation with $\epsilon$ held fixed at the value corresponding to the one between times $t_1$ and $t_2$, we observe a general nucleation-growth process. In this scenario, we can employ the Avrami equation with the time-independent  nucleation rate $\Gamma$ and the bubble wall velocity $v$,
\begin{eqnarray}\label{s2}
\mathfrak{f}_{\rm fixed}=1-\exp \left(-\frac12  v (\epsilon_{\rm fixed}) \Gamma (\epsilon_{\rm fixed})  t^2\right).
\end{eqnarray}

\begin{figure}
{%
\includegraphics[clip,width=0.4\columnwidth]{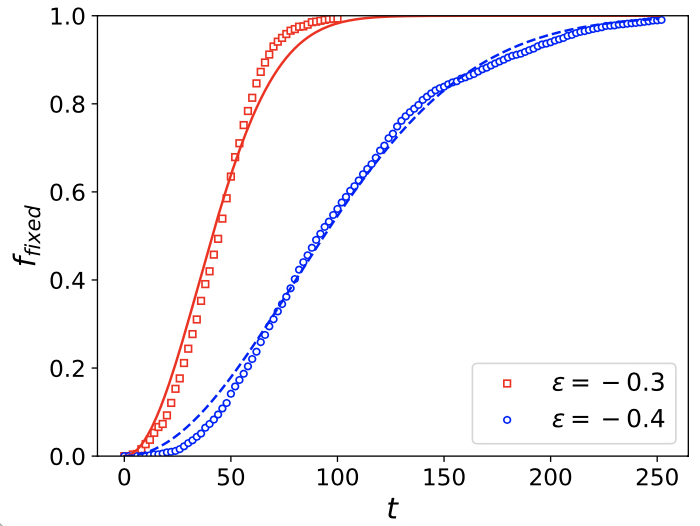} 
}
\caption{Time evolution of the  fraction of space covered by the new phase due to nucleations for $c=1$ and $\theta=0.01$. $\epsilon=-0.3$ (red squares) and $\epsilon=-0.4$ (blue circles). The lines represent the best fit with function (\ref{s2}).} 
\label{figsupp}
\end{figure}

We numerically obtain the time evolution of the  fraction of space occupied by the new phase due to nucleation events for each $\epsilon$ as in  Fig. \ref{figsupp}. By fitting the results  with the function (\ref{s2})  across a range of  $\epsilon$ values, we can estimate the dependence of $v$ on $\epsilon$. The observation indicates that $v$ approximately increases linearly with $\epsilon$, such that $v(\epsilon) = 0.026\epsilon + 0.016$ for $c=1$ and $\theta=0.01$. For different values of $c$ and $\theta$, we repeated the procedure described above to derive the equation for $v(\epsilon)$ corresponding to each set of $c$ and $\theta$ values. 

\subsection*{Number of defects generated by nucleation-growth processes}

\begin{figure}
{%
\includegraphics[clip,width=0.4\columnwidth]{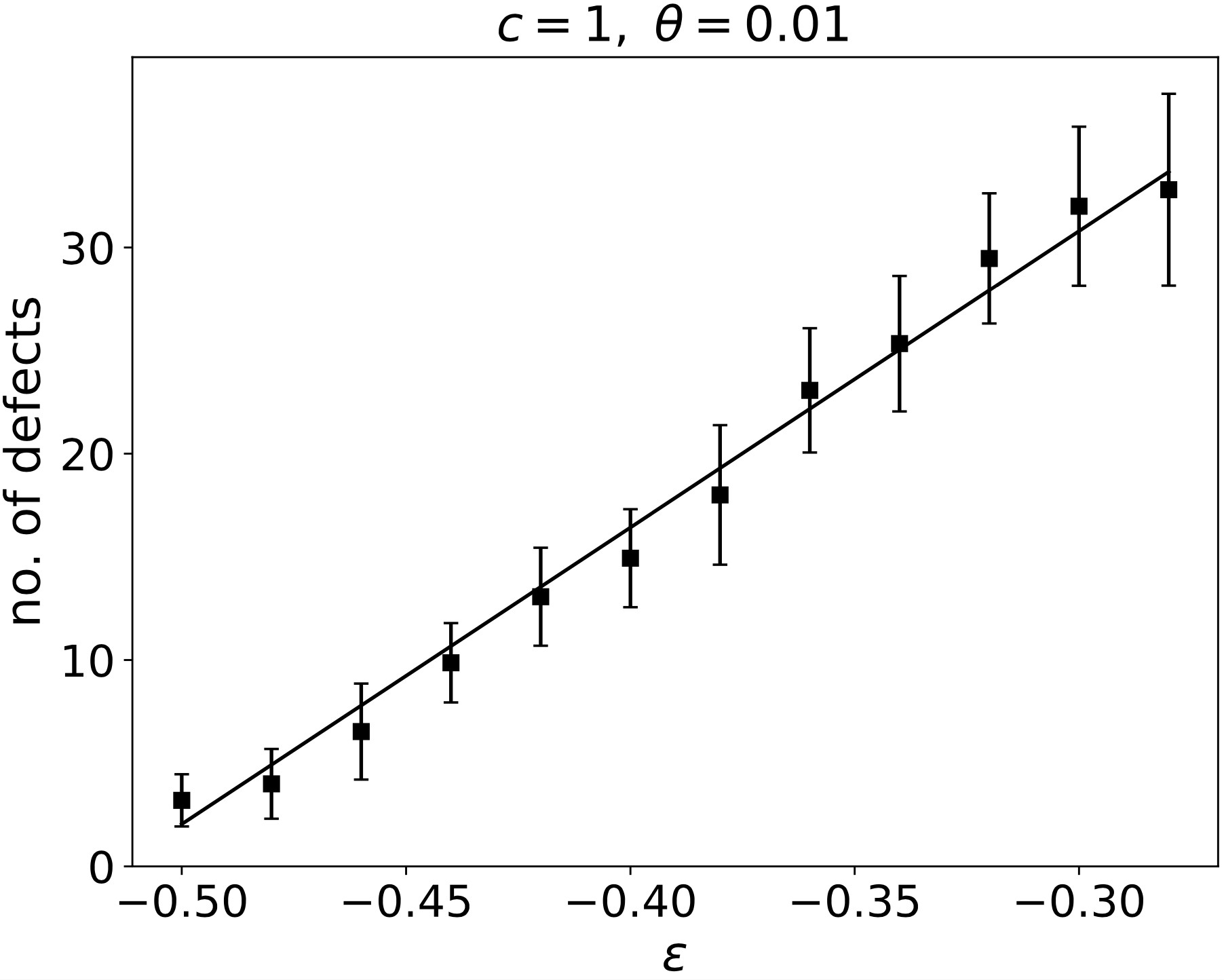} 
}
\caption{Number of defects as a function of $\epsilon$  solely generated by  nucleation-growth processes. The numerical results are represented by the squares, and the best fit is depicted as the line. $c=1$ and $\theta=0.01$.} 
\label{figsupp2}
\end{figure}

Fig. \ref{figsupp2}  was obtained by performing numerical simulations of nucleation-growth processes 15 times for each $\epsilon$. It can be seen that the number of defects increases as $\epsilon$ approaches 0, and the linear fit closely matches the numerical results depicted by the squares. We  use the  equation representing the best fit,  $n_{nuc} (\epsilon)=144\epsilon +74$ for $c=1$, $\theta=0.01$ in the  paper. For different values of $c$ and $\theta$, we repeated  same numerical procedure  and derived the  equation for $n_{nuc}$ for each set of $c$ and $\theta$ values.

Here $n_{nuc}(\epsilon)$ represents the number of defects created by nucleation for fixed $\epsilon$. Since $\epsilon$ changes over time during the phase transition, $n_{nuc}(\epsilon)$ also depends on time. Therefore, the number of defects  created throughout the phase transition can be approximated by
\begin{eqnarray}
\mbox{\# of defects created by nucleation} \approx \sum_{i=1}^{N} n_{nuc} (\epsilon (t_1+i\Delta t))\Delta \mathfrak{f}
\end{eqnarray}
where  we assumed that the space gets covered by new phase by $\Delta \mathfrak{f}$ for each time step $\Delta t$. The fraction of space occupied by the new phase becomes $\mathfrak{f}$ at time $t=t_1+N\Delta t$. Since $\Delta \mathfrak{f} = \mathfrak{f}/N$, we have
\begin{eqnarray}
\mbox{\# of defects created by nucleation} \approx \sum_{i=1}^{N} n_{nuc} (\epsilon (t_1+i\Delta t))\mathfrak{f}/N \approx n_{nuc} (\epsilon (t_1+N\Delta t/2))\mathfrak{f}= \mathfrak{f} n_{nuc} (\epsilon^{*})
\end{eqnarray}
where $\epsilon^{*}=\epsilon (t_1+N\Delta t/2)$ represents the value of $\epsilon$ at the time when the fraction of space occupied by the new phase reaches half of $\mathfrak{f}$. This outcome corresponds to  the first term of Eq. (10)  in the paper. Here we used the fact that $n_{nuc} (\epsilon)$ can be approximated by a linear function of  $\epsilon$.  Although $\Delta \mathfrak{f}$ is  time-dependent in general, it approaches 0 as $\mathfrak{f}$ approaches 0 and 1, while  it can be typically represented by a smooth positive function with a single peak  near the midpoint when $0<\mathfrak{f}<1$. Due to this behavior of $\Delta \mathfrak{f}$, the approximation of the number of defects produced by nucleation throughout the phase transition via  $n_{nuc} (\epsilon^{*})$  proves to be effective.

\end{document}